\documentclass{osa-article}
\journal{oe}

\articletype{Research Article}

\begin{document}

\title{Raman signal extraction from CARS spectra using a learned-matrix representation of the discrete Hilbert transform}

\author{Charles H. Camp Jr.}

\address{Biosystems and Biomaterials Division\\ National Institute of Standards and Technology\\100 Bureau Dr., Gaithersburg, MD 20899}

\email{charles.camp@nist.gov} 



\begin{abstract*}
Removing distortions in coherent anti-Stokes Raman scattering (CARS) spectra due to interference with the nonresonant background (NRB) is vital for quantitative analysis. Popular computational approaches, the Kramers-Kronig relation and the maximum entropy method, have demonstrated success but may generate significant errors due to peaks that extend in any part beyond the recording window. In this work, we present a learned matrix approach to the discrete Hilbert transform that is easy to implement, fast, and dramatically improves accuracy of Raman retrieval using the Kramers-Kronig approach.
\end{abstract*}

\section{Introduction}
Coherent anti-Stokes Raman scattering (CARS) microscopy and spectroscopy have long been proposed as a method for label-free chemical analysis and imaging at orders of magnitude higher speeds than traditional Raman micro/spectroscopy and at higher spatial resolution than afforded by infrared absorption\cite{Zumbusch1999, Cheng2002, Wurpel2002}. CARS uses coherent stimulation of molecular vibrations that enables both more efficient scattering of the probe beam and non-isotropic radiation of the output signal, which dramatically improves collection efficiency\cite{Petrov2007}. Additionally, the co-generation of a so-called nonresonant background (NRB) acts as a stable homodyne amplifier, greatly amplifying the signal\cite{Muller2007, Camp2014} --- without which CARS would often have limited advantage over spontaneous Raman\cite{Cui2009}. Though beneficial, the NRB distorts spectral peak shapes and contributes to the CARS signal having a quadratic polynomial dependence on analyte concentration, potentially spoiling quantitative analysis capabilities. To combat such challenges, optical methods to reduce the NRB were originally proposed; though, they reduced the overall CARS signal\cite{Muller2007,Cheng2001,Ganikhanov2006,Pestov2007}. Later research revealed that computational methods, based on electromagnetic\cite{Liu2009} or information theory\cite{Vartiainen1992}, could remove the distortion from the NRB without a reduction in signal strength\cite{Cicerone2012}. Additionally, these computational methods result in a signal that is linear with analyte concentration, enabling quantitative analysis of biological, chemical, and polymer samples\cite{Masia2013,Rinia2007,Karuna2016, Schafer2008, Lee2011}. Ideally, these methods require independent measurements of NRB spectra, which is not currently possible, leading to spectroscopic errors and distortions\cite{Camp2016}; though, experimental\cite{Karuna2016} and computational\cite{Camp2016} mitigations have been developed. Even with a known NRB, these methods suffer from artifacts when spectroscopic peaks (and ``wings"/``tails") extend beyond the recorded spectroscopic window. These artifacts distort spectral lineshapes and can even generate new features with similar intensities to the analyte signals. In this work, we present a new method of performing the Hilbert transform for the Kramers-Kronig (KK) relation method of extracting Raman spectral features from CARS spectra. The Learned Discrete Hilbert Transform (LeDHT) uses a synthetically generated set of single-peak training data (with known Hilbert transforms) to learn a matrix representation of the Hilbert transform. After learning (``training" in machine learning parlance), arbitrary new input spectra can be transformed via matrix multiplication, which is computationally fast and efficient. Furthermore, even with synthetic training data, the LeDHT matrix is usable on experimental data with superior performance as will be demonstrated with CARS spectra of glycerol.

\section{Theory}
\subsection{CARS spectroscopy and Raman signal extraction using the Kramers-Kronig relation}
CARS is a third-order nonlinear optical mechanism in which the difference frequency between a ``pump" photon and a ``Stokes" photon is capable of stimulating a Raman (vibrational or rotational) transition from which a ``probe" photon can inelastically scatter\cite{Gomez1996, Cheng2002, camp_jr_chemically_2015, Muller2007, Camp2014, Potma2013}. From a macroscopic and scalar perspective, the generated CARS spectral intensity, $I_{\mathrm{CARS}}(\omega)$, may be described as\cite{Camp2016}:
\begin{align}
    I_{\mathrm{CARS}}(\omega) ~\propto&~ \left| \left \{ \left [E_\mathrm{S}(\omega) \star E_\mathrm{p}(\omega) \right ]\chi^{(3)}(\omega) \right\} \ast E_{\mathrm{pr}}(\omega)\right|^2 ~\approx~ \left | \tilde{C}_\mathrm{st}(\omega)\right|^2 \left |\tilde{\chi}^{(3)}(\omega)\right |^2\\
    \chi^{(3)}(\omega) ~=&~ \chi_\mathrm{r}(\omega) + \chi_\mathrm{nr}(\omega)\\
    \tilde{C}_\mathrm{st}(\omega) ~\equiv&~ \frac{[E_\mathrm{S}(\omega) \star E_\mathrm{p}(\omega)] \ast E_{\mathrm{pr}}(\omega)}{\int E_{\mathrm{pr}}(\omega) d\omega}\\
    \tilde{\chi}^{(3)}(\omega) ~\equiv&~ \chi^{(3)}(\omega) \ast E_{\mathrm{pr}}(\omega),
\end{align}
where $E_\mathrm{p}$, $E_\mathrm{S}$, and $E_{\mathrm{pr}}$ are the pump, Stokes, and probe spectral fields (in frequency $\omega$), respectively; $\chi^{(3)}$ is the third-order nonlinear susceptibility; and $\star$ and $\ast$ are cross-correlation and convolution operators, respectively. The third-order nonlinear susceptibility can be separated into a summation of vibrationally (or rotationally) resonant ($\chi_\mathrm{r}$) and nonresonant ($\chi_\mathrm{nr}$) components, the latter of which is responsible for the aforementioned NRB. For simplicity, we have defined $\tilde{C}_\mathrm{st}$ that describes the effective stimulation intensity and bandwidth, and $\tilde{\chi}^{(3)}$ is the nonlinear susceptibility with intensity and spectral resolution as set by the probe source. Though neglected here, one could also account for detector and filter spectral response characteristics through modification of this term as well.

Spontaneous Raman spectroscopy records peaked spectra that are proportional to $\Im\{\chi_\mathrm{r}(\omega)\}$\cite{Tolles1977}, but CARS spectra are proportional to $|\chi^{(3)}(\omega)|^2$; thus, computational or experimental approaches need be taken to ascertain the phase of $\chi^{(3)}$. The two primary computational approaches (also known as ``phase-retrieval" methods) are the maximum entropy method (MEM)\cite{Vartiainen1992} and the Kramers-Kronig relation (KK)\cite{Liu2009}, which have been shown to be functionally equivalent\cite{Cicerone2012}. Mathematically, using the KK relation as derived in  \cite{Camp2016}, the retrieved nonlinear susceptibility (normalized to the NRB) is:
\begin{align}
    \frac{\chi_\mathrm{r}(\omega)}{\chi_\mathrm{nr}(\omega)} ~=&~ \underbrace{\sqrt{\frac{I_{\mathrm{CARS}}(\omega)}{I_{\mathrm{NRB}}(\omega)}}}_{A(\omega)} \exp \left [i~  \underbrace{\mathcal{H} \left \{\frac{1}{2}\ln \frac{I_{\mathrm{CARS}}(\omega)}{I_{\mathrm{NRB}}(\omega)} \right\}}_{\phi(\omega)}\right],\label{Eq:Raman_Extract}
\end{align}
where $I_{\mathrm{NRB}}(\omega)$ is the spectrum of the NRB, $\mathcal{H}$ is a Hilbert transform, $\phi$ is the phase, and $A$ is the amplitude (unitless ratio). The retrieved $\chi_r$ is normalized by $\chi_\mathrm{nr}$ in order to remove $\tilde{C}_\mathrm{st}$ (in both MEM and KK methods)--- a quantity that has not been shown to be independently measurable but may be assumed to be constant between different CARS measurements\cite{Masia2013}. Without loss of generality, we will treat the NRB as measurable in the der of this manuscript.

\subsection{Hilbert transform}
The Hilbert transform is a widely used linear operator implemented to extract analytic representations of signals from real-valued measurements. Simply put, causal systems may be represented by complex-valued functions in the frequency domain with a defined relationship between the real and complex components\cite{Toll1956}. The Hilbert transform lets one calculate the imaginary (real) portion of this function when only the real (imaginary) portion are directly measurable. For example, a molecular system with inhomogeneous broadening may present itself as a Gaussian lineshape through a spectroscopy that measures the real (imaginary) part of the system response, and its imaginary (real) part -- with analogous information -- would be a Dawson lineshape. There is also a similar relationship for when one only measures the magnitude (phase) under certain conditions\cite{smith_dispersion_1977}. The Hilbert transform is used across a myriad of application spaces such as vibrational analysis\cite{Fang2009,Feldman2011,Luo2009}, biomedical signal processing (e.g., cochlear implants\cite{Nie2006}, neural\cite{Tass1998,LeVanQuyen2001} and cardiac monitoring\cite{Benitez2001}), and spectroscopies (e.g., optical\cite{Liu2009, Camp2016, Robinson1952, Roessler1965}, terahertz\cite{THZSpectroscopy}, impedance\cite{Agarwal2019}, dielectric\cite{Dielectric}, magnetic resonance\cite{NMR, NMR2}), where it is often referred to as a Kramers-Kronig ``transform" or ``relation". Further, the Hilbert transform is an integral part of the analysis tool known as the Hilbert-Huang transform: a NASA-developed method most frequently applied to geophysical analyses\cite{Huang1998,Huang2008}. A significant hurdle to the accurate use of the Hilbert transform, however, is that the discrete implementation may deviate significantly from the continuous form.

The Hilbert transform, $\mathcal{H}$, of a function $f(x)$ on the real line is defined as
\begin{equation}
    \mathcal{H}\{f\}(x) = \frac{\mathcal{P}}{\pi}\int_{-\infty}^\infty \frac{f(x')}{x - x'}dx' = f(x) \ast \frac{\mathcal{P}}{\pi x},\label{Eq:Hilbert}
\end{equation}
where $\ast$ is a convolution and $\mathcal{P}$ is the Cauchy principle value. Though this may be solved analytically for known functions\cite{Poularikas1999}, under more experimentally relevant conditions with discrete data there are a myriad of proposed approaches to the discrete Hilbert transform (DHT). Broadly speaking, these methods approximate the integral in Eq. \ref{Eq:Hilbert} to a more tractable form\cite{Kress1970, Weideman1995}, model the input signal as a linear superposition of functions with analytically known Hilbert transforms\cite{Micchelli2013,Stenger1976,Zhou2009}, or solve the equivalent problems using the discrete Fourier transform\cite{Marple1999,Henrici1986}, which is a particular case of modeling the signal with known functions (sin and cos).

Using the Fourier transform is the most common approach in-so-much that it is broadly available in such computational software as GNU Octave, MATLAB, SciPy, and SAS. The aforementioned software packages use a fast-Fourier transform (FFT)-based method developed by Marple\cite{Marple1999} that uses symmetry properties of the discrete Fourier transform (DFT) to calculate the DHT. Another Fourier transform approach by Henrici\cite{Henrici1986}, which is implemented in LabView (and in previous work of the author\cite{Camp2016}), uses the (continuous) Fourier transformed version of Eq. \ref{Eq:Hilbert} as the starting point for the computation with a FFT. The mathematical forms and the equivalency of these two methods are derived in the Supplemental Document subsection 1.A.

\subsection{Errors in the DHT}
Significant challenges to accurate DHT computation arise in the case of signals that are non-bandlimited, non-compactly supported, and non-periodic --- ``general inputs" per  \cite{DFT} --- such as is common in spectroscopies. The DHT on signals with these characteristics may have dramatic errors near the window edge --- ``end-"/``edge-effect" errors; though in fact, the error exists across the entire signal. These errors, as will be demonstrated, may cause significant errors in quantitative analysis of peak heights and relative ratios and may even generate new features.

Alternative DHT algorithms beyond Henrici and Marple primarily focus on computational efficiency and/or accuracy away from the window-edge; though, some methods show improvements for window-centered signals\cite{Bilato2014, Abd-el-Malek2020}. To address window-edge errors, padding schema have been proposed\cite{Huang2003} as well as analytical\cite{Meissner2012} and machine learning methods\cite{Cheng2007,Deng2001} to artificially extend the input signal as to push the edge-effect errors further from the original data. These methods, however, do not stop the generation of the edge effect errors, but rather move it further away from the window center to an artificially larger window. This, in some circumstances may be adequate, but as previously mentioned the edge effect errors permeate the entire window.

The aforementioned errors arise from the discretization and truncation of the input data. For example, the use of the DFT/FFT to effectively perform a circular convolution would see a difference in values from each edge of the signal window as a discontinuity. Similarly, a sudden change in derivative between the edges of the window would act as a form of discontinuity that would ultimately affect the correct calculation (related and bounded by the Paley-Wiener theorem\cite{DFT}). Two properties of the DHT that are derived below (the second being previously unpublished to our knowledge), provide a means to calculate the total error of using the DHT (supporting derivations are found within the Section 1 of the Supplemental Document):
\begin{enumerate}
    \item The mean of the DHT of a signal is zero
    \item The variance (and standard deviation) of the input signal is the same as the DHT of the signal.
\end{enumerate}
Using the mathematical definition of the discrete Fourier transform (DFT) from  \cite{DFT} (see also the Supplemental Document Section 1), the mean of the DHT of a length $N$ signal, $f[n] \in \mathbb{R}^N$, may be shown to always be zero:
\begin{align}
    \left<\mathcal{H}\{f\}[n]\right> ~=&~ \frac{1}{N}\sum_n \mathcal{H}\{f\}[n] = \mathcal{F}\{H\{f\}\}[0] = -i~ \text{sgn}[0]F[0] = 0,
\end{align}
where $\mathcal{F}$ is the DFT, $\mathcal{H}$ is the DHT, $F[k]$ is the Fourier-transformed signal at frequency $k \in \mathbb{Z}$, and `sgn' is the sign/signum function. For this derivation we have used the Henrici version of the DHT (see the Supplemental Document section 1.A).

Next, we show that the variance of a signal is the same as that of the DHT of the signal, which is not necessarily true for the continuous Hilbert transform. First, using Parseval's theorem:
\begin{align}
    \frac{1}{N}\sum_n \left | \mathcal{H}\{f\}[n]\right|^2 ~=&~ \sum_k \left | \mathcal{F}\{\mathcal{H}\{f\}\}[k]\right|^2 ~=~ \sum_k (\vec{1}-\delta[k])\left|\mathcal{F}\{f\}[k]\}\right|^2 \nonumber\\~=&~ \frac{1}{N}\sum_n \left|f[n]|^2 - |F[0]\right|^2\nonumber\\
    ~=&~ \frac{1}{N}\sum_n \left|f[n]\right|^2 - \left|\frac{1}{N}\sum_n f[n]\right|^2 ~=~ \sigma_f^2,
\end{align}
where $\vec{1}$ is a vector of 1's, $\delta$ is the unit impulse response (or unit sample function), and $\sigma_f$ is the standard deviation ($\sigma_f^2$ is the variance) of the signal $f[n]$. We have used the derivation of the Fourier transform of the Hilbert transformed signal that is derived in the Supplemental Document section 1.B. We can use this finding to analyze the variance of the DHT-transformed signal:
\begin{align}
    \sigma_{\mathcal{H}\{f\}}^2 ~=&~ \frac{1}{N}\sum_n \left|\mathcal{H}\{f\}[n]\right|^2 - \underbrace{\left|\frac{1}{N}\sum_n \mathcal{H}\{f\}[n]\right|^2}_{=~0}~=~\sigma_f^2,
\end{align}
where the second term is zero owing to first property of the DHT that was derived. This conservation of variance between the original signal and the DHT-transformed is not found when using analytically solved Hilbert transformed signals and is an artefact.

Though not provided in this manuscript, a detailed analysis of the DHT through analyzing bounds of decay relations in the DFT/FFT (e.g., see Chapter 6.7 in  \cite{DFT}) can provide insights and bounds onto the shape of the errors for known functions.

\subsection{LeDHT}
\begin{figure}[htbp]
\centering\includegraphics[width=5.25in]{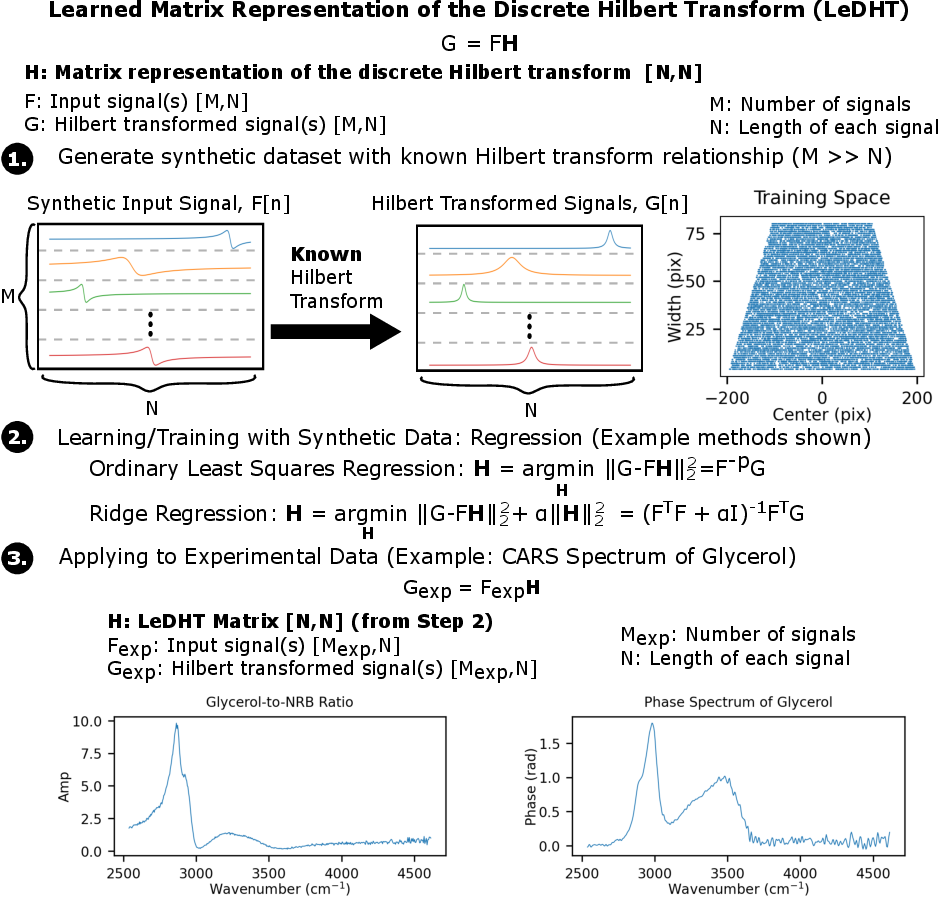}
\caption{Infographic describing the three steps of implementing the LeDHT: generation of synthetic training data, learning an optimal matrix form of the Hilbert transform, and application to new spectra via matrix multiplication. Note: Scatter plot points sub-sampled (5x) for visual clarity.}
\label{Fig:Infographic_smaller}
\end{figure}

As described above, the DHT has properties that adversely affect its accuracy --- especially in the common case of the spectrum extending to or beyond the recording window edge. Additionally, the DHT has fundamental properties that are aberrant when compared to the continuous Hilbert transform. Thus the aim of the LeDHT is to calculate the Hilbert transform on discrete data (spectra) with significantly fewer errors than the DHT. For CARS spectra, this means more accurate phase retrieval and Raman signal extraction than conventional methods. At its core, the LeDHT is a square matrix, $\mathbf{H} \in \mathbb{R}^{N\times N}$, which applies the transform via matrix multiplication: $G=F\mathbf{H}$, where $F$ is a matrix of input spectra and $G$ are the calculated Hilbert transforms. Both $F$ and $G$ $\in \mathbb{R}^{M\times N}$, where $M$ are the number of spectra and $N$ is the length of each spectrum. This matrix approach evolved from the fact that linear discrete transforms may be described by  matrix formulations \cite{riley_mathematical_2002} (e.g., Hilbert\cite{Burris1975, Dutta1976} and Fourier\cite{Kahaner1970, DFT} transforms). These matrix forms, however, are equivalent and generate the same errors and distortions. In the LeDHT, though, the matrix is not prescribed but rather learned through linear regression (least-squares) of synthetic training data with known [continuous] Hilbert transforms. This matrix form may be learned once for a given spectroscopic system and reused indefinitely -- assuming one's choice of lineshape(s) and training width ranges and center ranges are still optimal for the sample and optical systems. The three steps to training and applying the LeDHT are described pictographically in Fig. \ref{Fig:Infographic_smaller} and will be elaborated on below. 

The first step to implementing the LeDHT is the generation of synthetic training spectra. These $M$ spectra will be used to learn the square LeDHT matrix, $\mathbf{H} \in \mathbb{R}^{N\times N}$ ($M \gg N$), that describes the relationship between a spectrum of length $N$ and its Hilbert transform (See step 1 of Fig. \ref{Fig:Infographic_smaller}). The linear nature of the LeDHT offers significant benefits when generating synthetic data to span the available training space (i.e., possible lineshape parameter values). Of the most consequential benefits is that each training spectrum need only a single peak as more complex spectra can be composed of a linear superposition of single-peak spectra. This greatly reduces the number of possible unique spectra for training when compared to nonlinear models such as neural network-based approaches\cite{Valensise2020,Houhou2020} that require full multi-peak spectra. As an example, consider the use of Gaussian lineshapes with corresponding scaled ($2\pi^{-1/2}$) Dawson functions for the Hilbert transforms. These lineshapes have 2 free parameters: center position and a width parameter. Assuming $n \in \mathbb{Z}^N$, center position is an integer value on this line, the width is a positive integer value that also enables at least 3 samples per full-width at half-maximum (FWHM), and the maximum FWHM is $N$, the number of unique peaks is $(N-1)(N/2 -1)~=~N^2/2 - 3N/2 + 1$. For a 1000-point signal, 498\thinspace501 unique lineshapes exist under these conditions. Assuming double-point precision, the complete training dataset would be under 4 GB. This is dramatically lower than if full, multi-peak spectra were needed, such as with neural network approaches\cite{Valensise2020,Houhou2020}, where there are $\approx 2.9\times10^{85}$ unique spectra for a 1000-point signal length with 1 to 15 possible peaks.

Generation of the synthetic training data has several considerations such as lineshape function and any constraints on training parameter space. From earlier experiments, we found that Dawson lineshapes, of which Gaussian lineshapes are the Hilbert transform, generated more optimal results than dispersive and Lorentzian lineshape pairs. These lineshape types have only two training space dimensions to consider: center position (frequency) and width. A potentially more general lineshape for future inquiry would be using the Faddeeva function in which real part is a Voigt function and the imaginary part is its Hilbert transform. From a practical perspective of this work, the three parameters (two related to width and one to spectral position) would have greatly expanded our training space, limited the fraction of which we could sample (limited by available memory of the computer). The second important consideration are any constraints on the training space parameters. In this work, the minimum peak width (standard deviation for a Gaussian) was 4 spectral pixels, and the maximum was 80 spectral pixels. Additionally, any generated peak must be at least half a FWHM from the window edge; thus, as peaks are centered closer to the window edge, the maximum allowable width is narrower. This gives the training space parameter map a trapezoidal quality as seen in Fig. \ref{Fig:Infographic_smaller}. Additionally, we limited spectral position and width values to integer values (unit of spectral pixels in $n$). A description of how the training data was algorithmically generated is provided within the Methods Section.

The second step in implementing the LeDHT is solving for the $\mathbf{H}$ matrix in $G=F\mathbf{H}$, where $F$ are the input training spectra and $G$ are the known Hilbert transforms. Generally speaking, there are numerous methods to solve for an optimal LeDHT matrix. In this work, we are using ordinary least squares (OLS) which minimizes the residual sum-of-squares: $\mathbf{H} = \mathrm{argmin}_\mathbf{H} ||G - F\mathbf{H}||^2$. Beneficially, OLS is the most common linear regression method; thus, there are numerous efficient algorithms for its calculation. We have also tested common ``regularized" least squares methods such as ridge and LASSO regression with success though with inferior results for the data within this manuscript. For different lineshapes and different spectroscopic conditions, these other methods could potentially be superior.

Finally, the LeDHT can be applied to new data via matrix multiplication, which is computationally inexpensive: $G_{exp} = F_{exp}\mathbf{H}$, where $F_{exp}$ is a newly measured input spectrum/spectra and $G_{exp}$ is the LeDHT Hilbert-transformed spectrum to be calculated. Ideally, the solved-for LeDHT matrix for a given spectroscopy system does not need to be re-learned between experiments and can be reused indefinitely, again, as long as the training dataset still optimally reflects the expected future input spectral lineshape characteristics.

\section{Methods and materials}
\subsection{Processing and software}
All simulated and experimental data processing was performed on a Dell Precision 7730 laptop with a 6-core i7-8850H processor at 2.6 GHz with 64 GB of RAM. All processing of simulated and experimental data was performed using Python 3.8.5, NumPy, SciPy\cite{SciPy}, scikit-learn\cite{scikit-learn}, Matplotlib, and in-house developed Python packages. The traditional DHT used in this manuscript is the Marple version as available in SciPy. The LeDHT method, several other implementations of DHT methods (and padding schema), and Jupyter Notebook examples have been made available as an open-source Python package, \textit{Hilbert}, at \cite{Github}.

\subsection{Generation of synthetic data}
Synthetic data is generated in a stochastic manner in which the free parameters are randomly selected and accepted based on whether it is a unique combination and is within given bounds. For the simulations, $N$ is 401 with $n$ in the range [-200, 200]. The generated training data is composed of Dawson function lineshapes and their Hilbert transforms: Gaussian lineshapes. The minimum acceptable Gaussian standard deviation is 4 $n$-pixels and the maximum is 80. Furthermore, the center location of Gaussian lineshapes is bounded [-200,200] (see Fig. \ref{Fig:Infographic_smaller} ``Training Space" subimage) but that for a given peak, the center position must be at least half a FWHM from the window edge. Under these conditions, 100\thinspace000 of these unique single-lineshape spectra are generated. For further stability and improved results, we augmented the data by repeating the dataset 3 times, adding white Gaussian noise (standard deviation of $1\times10^{-12}$) and adding a constant offset sampled from a normal distribution. Thus the total training data was 300\thinspace000 single-peak spectra.

An additional 100\thinspace000 single-peak spectra were generated as test set for validation. This validation was repeated 3 times, for a total of 300\thinspace000 spectra. These spectra were noiseless and did not have a DC offset as to represent the ideal, expected spectral case. Results with test data that did contain noise and DC offset provided similar results to those presented in the Results Section.

For the simulated multi-peak spectra experiments, the same training data was used as for the single-peak experiments, but the test data (i.e., multi-peak spectra) was generated from linear summations of Dawson function lineshapes (and their corresponding Hilbert transform Gaussian lineshapes). Each randomly generated spectrum was composed of 1 to 15 peaks (uniform sampling). Each peak had an amplitude that was uniformly sampled in the range of 1 to 2 (no units), a width that was uniformly sampled between 4 and 26 (integer valued), and a center position that needed to be at least half a FWHM from the window edge.

For working with the experimental BCARS spectra that are 250 to 460 n-pixels long, we generated 150\thinspace000 Dawson and Gaussian lineshapes limited to widths of 3 to 50 n-pixels. Next we replicated (3x) and augmented the data in a similar manner as previously described, adding white Gaussian noise (standard deviation $10^{-3}$), and a DC offset sampled from a normal distribution. This provided a total of 450\thinspace000 training spectra.

\subsection{CARS spectrometer}
In this work, spectra of neat glycerol (BioUltra, Sigma) were collected on a previously developed broadband CARS (BCARS) microscope that is described in detail elsewhere\cite{Camp2014}. The glycerol was mounted between a coverslip (VWR) and glass slide (Ward's) separated by an adhesive well (GraceBio). Each spectrum was collected with a 3.5 ms dwell time and with average power of the probe and supercontinuum of 20.1 mW and 11.2 mW, respectively. Spectra of the coverslip were collected and used for normalization --- a process that effectively suppresses a co-generated, coherent nonresonant background\cite{Camp2014,Camp2016}. Additionally, dark spectra (250) were collected under the same conditions except with the laser sources not temporally overlapped (but still illuminating the sample). The glycerol spectrum shown in this work is an average of 97 individual spectra.

\section{Results}
\subsection{Simulated}
\begin{figure}[htbp]
\centering
\includegraphics[width=5.25in]{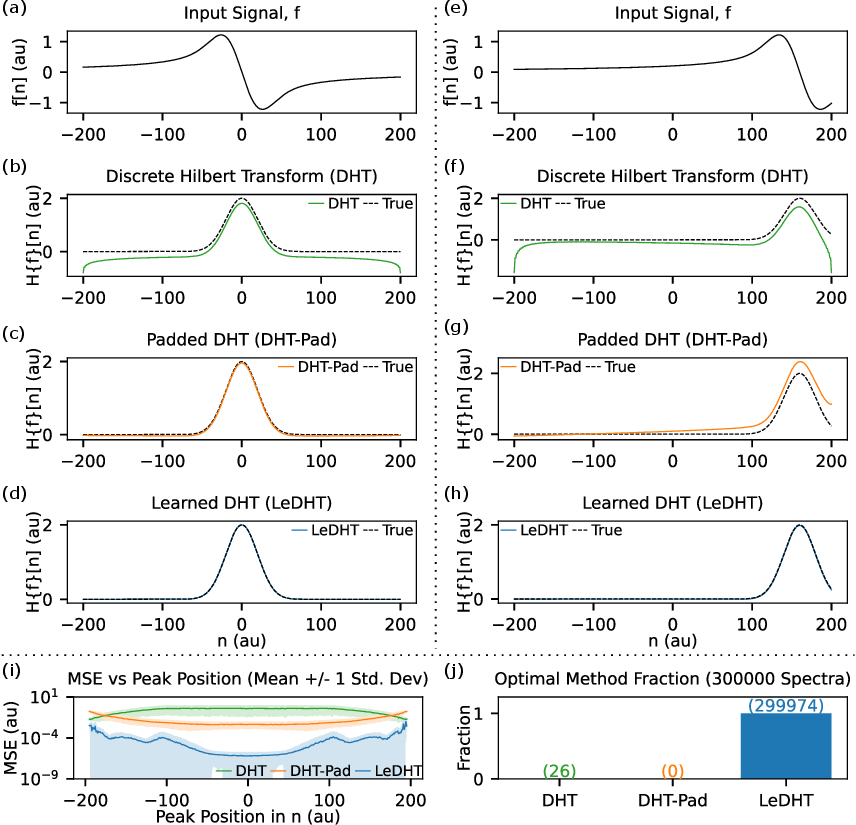}
\caption{(a,e) Example Dawson function input spectra. (b,f) DHT results showing significant errors. (c,g) Results using the DHT with extrema-value padding (DHT-Pad). (d,h) Results using the LeDHT. (i) Mean MSE ($\pm$ 1 standard deviation) calculated for the 300\thinspace000 stochastically generated test spectra as a function of peak center position. (j) Histogram of optimal method based on minimum MSE for test spectra dataset. Y-axis shows fraction, parenthetical values are number of spectra.}
\label{Fig:Synth_Peaks}
\end{figure}

As an initial demonstration, we will compare the DHT, the DHT with extrema-value padding (DHT-Pad) -- the most common padding scheme in the CARS community -- and the LeDHT. Figure \ref{Fig:Synth_Peaks}(a) is a representative Dawson function spectrum that is spectrally centered, and Figs. \ref{Fig:Synth_Peaks}(b-d) show the respective outputs of the DHT, DHT-Pad, and LeDHT. The DHT generates the most severe errors, with new downward-facing features at the spectroscopic window's edges. The magnitude of these errors are such that the mean of the DHT spectrum is zero as described previously in the Theory section. The DHT-Pad, by visual inspection, agrees with the true Hilbert transform with an apparent slight constant offset. Finally, the LeDHT appears to agree with the true Hilbert transform. Using mean squared error (MSE) as the error metric, the LeDHT is approximately 6 orders of magnitude better than the DHT-Pad and 8 orders of magnitude improvement over the DHT.

Next, we examine the results of using the same input Dawson spectrum but shifted closer to a window edge, as shown in Fig. \ref{Fig:Synth_Peaks}(e). Similar to before, the DHT produces erroneous downward facing features that are significant in amplitude as shown in Fig. \ref{Fig:Synth_Peaks}(f). Additionally, the DHT peak maximum value is approximately 20 \% less than the true value. With the spectral peak near the window edge, now the DHT-Pad is demonstrating significant inaccuracies. The peak maximum is significantly above the ideal case ($>20 \%$ at the peak maximum) and there is a clear distortion in the baseline. Figure \ref{Fig:Synth_Peaks}(h) again demonstrates the superior performance of the LeDHT. The MSE of the LeDHT is approximately 550 times lower than the DHT-Pad, and nearly 1\thinspace200 times lower than for the DHT.

To further examine the peak position-error relationship, Fig. \ref{Fig:Synth_Peaks}(i) shows the MSE (on a log scale) as a function of peak position from the 300\thinspace000 test data spectra. The LeDHT has the lowest mean MSE of all methods across all peak positions. The next best method is DHT-Pad, which outperforms DHT across the entire window save the edges. And finally, we calculated for each of the 300\thinspace000 test spectra, which method had the best performance as shown in Fig. \ref{Fig:Synth_Peaks}(j). The LeDHT had the best performance in 299\thinspace974 of 300\thinspace000 spectra ($\approx$ 99.99 \%), with the DHT having better MSE 26 times ($\approx$ 0.01 \%). All 26 results were from narrow peaks centered at the window edges, for which improved training data with some added emphasis on these cases would likely improve LeDHT performance. The DHT-Pad was never the optimal choice, which indicates that the LeDHT always outperforms the DHT-Pad. Section 2.A of the Supplemental Document contains the converse scenario in which input spectra are single Gaussian peaks and the resultant Hilbert transforms (Dawson functions). Like the results presented here, the LeDHT is similarly superior. Section 2.C of the Supplemental Document also demonstrates using the Gaussian training set above on test data that is Lorentzian lineshapes. Surprisingly, the LeDHT still outperforms the DHT and DHT-Pad on 300\thinspace000 test spectra.

\begin{figure}[htbp]
\centering
\includegraphics[width=5.25in]{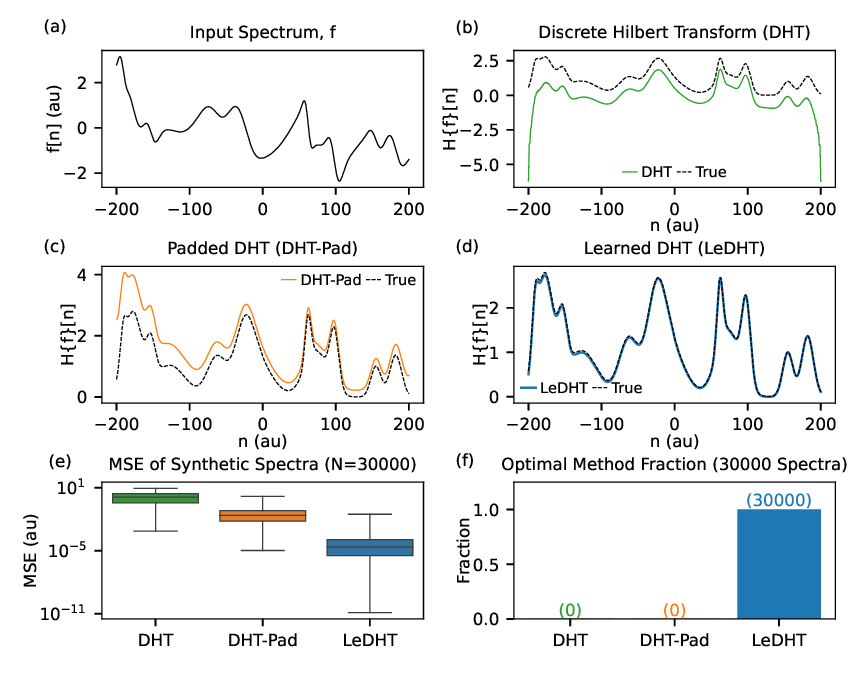}
\caption{(a) Example synthetic spectrum composed of 13 Dawson functions. (b,c,d) The true Hilbert transform and the calculated Hilbert transforms using DHT, the DHT applied to extrema-value padded spectra (DHT-Pad), and the LeDHT, respectively. (e) Box plots of MSE for each method. Note the whiskers displays the extreme range of values. (f) Histogram of the optimal method (based on minimum MSE) for 30\thinspace000 test spectra. Y-axis shows fraction, parenthetical values are number of spectra.}
\label{Fig:Synth_multi-peak_Spectra}
\end{figure}

Next we analyzed the DHT, DHT-Pad, and LeDHT performance when applied to synthetically generated multi-peak spectra. 30\thinspace000 spectra were generated stochastically, each with 1 to 15 peaks. Figure \ref{Fig:Synth_multi-peak_Spectra} (a) shows an example single spectrum composed of a linear combination of 13 Dawson function lineshapes. Figures \ref{Fig:Synth_multi-peak_Spectra}(b-d) shows the corresponding transforms. Looking at the DHT, again one can see downward sloping features at the window edge that are larger than any real peak (though negative) with over 1000 \% relative error at the window edges (over 5800 \% at the right side). Additionally, one can see that the entire DHT spectrum is offset below ideal outcome leading to a minimum relative error of $\approx$ 29.5 \%. Similarly, the DHT-Pad spectrum is entirely shifted above the ideal spectrum with a minimum relative error of $\approx$ 8.5 \%. The DHT-Pad also shows amplitude errors that become worse at the window edges: greater than 300 \% relative error at the left edge and 550 \% at the right edge. In contrast, the LeDHT is significantly less biased with a minimum relative error of $\approx$ 0.04 \%. The window edge extrema relative errors were 13 \% (left) and 0.15 \% (right). Figure \ref{Fig:Synth_multi-peak_Spectra}(e) shows box plots of MSE generated by the three methods. The LeDHT median MSE is $\approx$ 4.5 orders of magnitude less than that of the DHT, and approximately 3 orders of magnitude less than the DHT-Pad. Figure \ref{Fig:Synth_multi-peak_Spectra}(f) further affirms the superiority of LeDHT, identifying that the LeDHT outperformed, based on MSE, the other methods in all 30\thinspace000 test spectra. Section 2.B of the Supplemental Document contains the converse scenario in which input multi-peak spectra are composed of Gaussian lineshapes and the resultant Hilbert transforms, displaying similarly superior results.

As a final note, we compared the computational time of the DHT, DHT-Pad, and LeDHT methods. We do not assume that any of our algorithms are optimal; thus, there is likely much room for improvement. For 100\thinspace000 single-peak spectra, the DHT and DHT-Pad took approximately 4 seconds and 15 seconds, respectively over 3 trials. The LeDHT, took 10 seconds to train on 300\thinspace000 spectra, and only 0.4 seconds to 0.5 seconds to calculate 100\thinspace000 test spectra. And as mentioned previously, the LeDHT can be trained once and reused in future experiments; thus, the LeDHT was approximately an order of magnitude faster than the DHT when already trained, and over 30 times faster than DHT-Pad.

\subsection{Experimental CARS spectra}
Finally, we applied the LeDHT to experimental CARS spectra of neat glycerol, varying the proximity of the spectrum to the window edge to demonstrate the superior repeatability and stability of the LeDHT method. Though this scenario may seem artificial, for broadband ($>$3500 cm$^{-1}$) or low-frequency ($<$500 cm$^{-1}$) applications, peaks extending to the window edge become probable (especially in chemical and biological samples). The Raman phase ($\phi$, Eq. \ref{Eq:Raman_Extract}) and Raman:NRB ratio  ($\chi_R/\chi_{NR}$, Eq. \ref{Eq:Raman_Extract}) were calculated as described in the Theory and Methods section. These quantities were calculated at different signal bandwidths: from $\approx$ 2080 cm$^{-1}$ down to $\approx$ 1020 cm$^{-1}$ in 23 increments. The training data was composed of 450\thinspace000 Gaussian lineshapes augmented with noise and DC bias as described in the Methods section. 

\begin{figure}[htbp]
\centering
\includegraphics[width=5.25in]{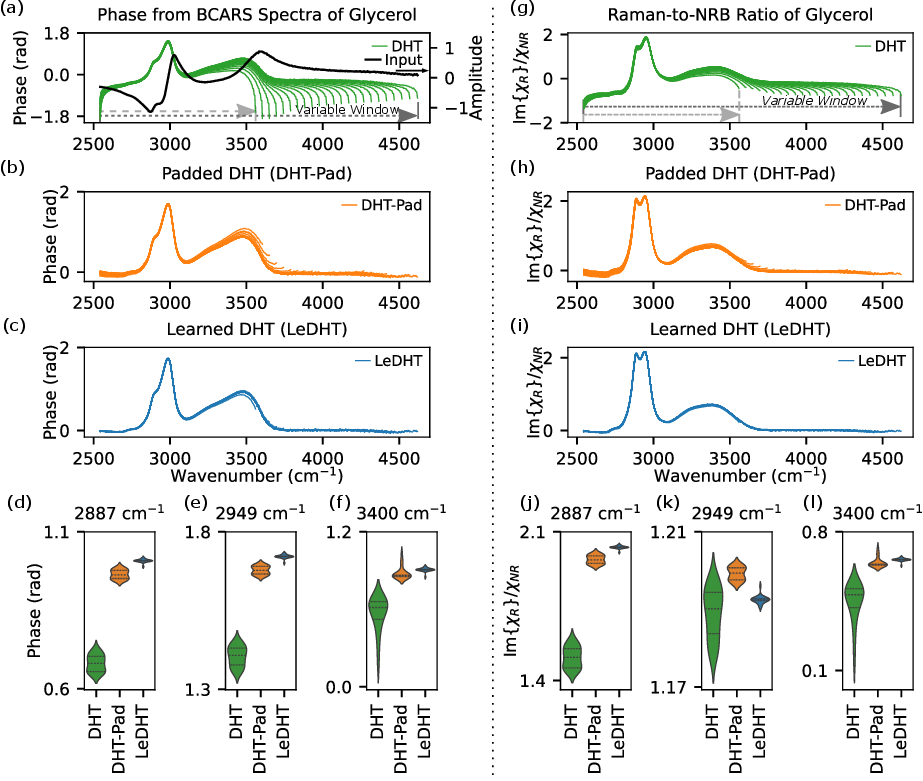}
\caption{(a) Black line associated with right y-axis: negative half natural log of experimental BCARS spectrum from glycerol normalized to coverslip, which is the input to the Hilbert transform for phase calculations. Green lines: DHT-calculated phases when the input is truncated at different spectral widths. (b) Phase calculations using extrema-valued padded spectra into the DHT (DHT-Pad). (c) Phase calculation using the LeDHT. (d-f) Violin plots of the phase magnitude for three prominent peaks. (g-i) Raman-to-NRB ratio calculations using DHT, DHT-Pad, and LeDHT methods, respectively. (j-l) Violin plots of the extracted Raman-to-NRB ratio for prominent peaks.}
\label{Fig:Experimental}
\end{figure}

Figure \ref{Fig:Experimental}(a) shows the input to the Hilbert transforms, $-0.5 \text{ ln} (I_{\mathrm{CARS}}/I_{\mathrm{NRB}})$, where the NRB was recorded from coverslip glass. Note the negative sign when compared to Eq. \ref{Eq:Raman_Extract}, due to our experimental data vectors being oriented with decreasing $\omega$ vectors (i.e., the $\omega$ vector is numerically high-valued to low-valued). Figures \ref{Fig:Experimental}(a) and (g) show the Raman phase and Raman:NRB ratio, respectively, as extracted using the DHT. As expected, the DHT generates large negative features at the window extrema. Figures \ref{Fig:Experimental}(b) and (h) show the DHT-Pad results that are by inspection significantly improved. One can see, though, that as the bandwidth of the spectrum is changed, the resulting Hilbert transform is modulated in amplitude, especially at the baseline near 2600 cm$^{-1}$ and at the OH-stretch maximum near 3400 cm$^{-1}$. Conversely, the peak at 2949 cm${-1}$ is stable. A common approach in CARS spectroscopy is to set the baseline to 0 at the window edge. Doing so, in this case, at approximately 2600 cm$^{-1}$ would adversely cause the stable amplitude peak at 2949 cm$^{-1}$ to modulate. Finally, Figs. \ref{Fig:Experimental}(c) and (i) show the LeDHT results that are significantly more stable with window width and only appear to fluctuate mildly within the OH-stretch region. To qualify the stability of the LeDHT algorithm, Figs. \ref{Fig:Experimental}(d) and (j) present violin plots of the peak phase and Raman:NRB at 2887 cm$^{-1}$, 2949 cm$^{-1}$, and 3400 cm$^{-1}$ across the various window widths. In all six plots, the vertical extent of the LeDHT violin (indicating variation) is significantly smaller than for DHT and DHT-Pad. For example, for the 3400 cm$^{-1}$ peak, the mean $\pm$ standard deviation phase for the DHT is (0.50 $\pm$ 0.12) rad, (0.8 $\pm$ 0.04) rad for the DHT-Pad, and (0.83 $\pm$ 0.01) rad for the LeDHT. Thus in short, the LeDHT standard deviation is about 12 times less than the DHT and 4 times less than the DHT-Pad. For the Raman:NRB calculations, the LeDHT standard deviation is similarly 10 times less than the DHT and 3 times smaller than the DHT-Pad.

\section{Discussion and conclusion}
In this work, we have presented a learned matrix approach for calculating the Hilbert transform with superior performance compared to the DHT and a padded DHT schema. These demonstrations were performed on a variety of synthetic signals as well as experimental CARS spectra. And though the LeDHT demonstrated orders-of-magnitude improvement in MSE on the synthetic datasets and heightened stability with experimental data, there are still numerous opportunities for further gains and new application spaces to consider.

There are many extensions and improvements to LeDHT that could further enhance and expand the method. For example, the training data could be augmented to explicitly diminish noise. In this case, it may be advantageous to repeat the training dataset with random additions of noise solely on the input. Our initial attempts at this were successful though areas of low-signal and high-noise were affected in a similar manner to traditional [convolution] filtering methods (e.g., bandpass, Savitzky–Golay) with erroneous lineshape-like features being generated within the noise. Traditionally, we have had more success with separate processes of statistical denoising\cite{Camp2016}.

Another opportunity for improvement is methods to effectively train with smaller datasets, which in turn enables the generation of larger and larger training datasets with unique features. For examples, we have experimented with generating large training datasets and factorizing with singular value decomposition (SVD), which effectively generates a much smaller training dataset with each entry being more complex than a single lineshape. The 300\thinspace000 training spectra used for Figs. \ref{Fig:Synth_Peaks} and \ref{Fig:Synth_multi-peak_Spectra}, for example, could be reduced down to 401 (matching the size of $N$) -- a nearly 750x reduction. In the long term, this could enable a training dataset to be originally generated from numerous different lineshapes, potentially enabling a more universal LeDHT matrix that could be shared.

We are also exploring alternative regression/optimization methods for finding the LeDHT matrix. This matrix may be more optimal in terms of MSE, or it may be more optimal in the sense of having a symmetry (e.g., a Procrustes problem) that can be more easily human-interpreted or implemented in hardware. Thus far we have worked on the former problem: implementing Ridge and LASSO regression methods; though linear regression has performed most optimally.

And finally, we are looking to apply LeDHT to other signal types and spectroscopic modalities. Although CARS is a relatively niche technique, methods such as NMR\cite{NMR, NMR2}, dielectric\cite{Dielectric}, and terahertz\cite{THZSpectroscopy} spectroscopies have also demonstrated use of the DHT. 

In closing, we have developed LeDHT and demonstrated significantly improved MSE over the traditional DHT and with signal extrema padding. Our method is fast, stable, and improves opportunities for quantitative analysis of spectroscopic signals. Additionally, an open-source Python library for this method is freely available to enable community development and foster adoption.

\begin{backmatter}
\bmsection{Acknowledgements}
The author would like to thank the peer-reviewers for their effort and insightful comments. Additionally, the author would like to thank Young Jong Lee, Nancy Lin, Sheng Lin-Gibson, Donald Atha, John Henry Scott, and the internal review board (WERB) at NIST for their helpful reviews and recommendations towards the preparation of this manuscript.

\bmsection{Disclaimer}
Any mention of commercial products or services is for experimental clarity and does not signify an endorsement or recommendation by the National Institute of Standards and Technology. The authors declare no conflicts of interest.

\bmsection{Disclosures}
The authors declare no conflicts of interest.

\bmsection{Data availability}
The scripts, data, and library used in this manuscript are publicly available at \cite{Github}.

\bmsection{Supplemental document}
See Supplement 1 for supporting content.
\end{backmatter}

\bibliography{biblio}
\end{document}


\maketitle

\section{Mathematical derivations}
In this work, we are considering an odd-length signal $f[n]$ where $-(N-1)/2 \leq n \leq (N-1)/2$ and are using the following definitions for the DFT and iDFT per \cite{DFT}:
\begin{align}
    F[k] ~=&~ \mathcal{F}\{f\}[k] := \frac{1}{N}\sum_{n= -\frac{N-1}{2}}^{\frac{N-1}{2}}f[n] e^{-i\frac{2\pi}{N}kn}\\
    f[n] ~=&~ \mathcal{F}^{-1}\{F\}[n] := \sum_{k=-\frac{N-1}{2}}^{\frac{N-1}{2}}F[k] e^{i\frac{2\pi}{N}kn},
\end{align}

\subsection{DHT of Henrici and Marple are equivalent}\label{Appendix_Henrici_Marple}
For an odd-length signal the DHT implementations of Marple \cite{Marple1999} and Henrici \cite{Henrici1986} are:
\begin{align}
    \text{Marple}: \mathcal{H}\{f\}[n] = \text{Im}\left\{\mathcal{F}^{-1}\left.\begin{cases}
    0 & k < 0\\
    \mathcal{F}\{f\}[k] & k=0\\
    2\mathcal{F}\{f\}[k] & 1\leq k \leq \frac{N-1}{2}\\
    \end{cases}\right\}\right\}\label{Eq:Marple}\\
    \text{Henrici}: \mathcal{H}\{f\}[n] = \mathcal{F}^{-1}\{-i\, \text{sgn}[k] \mathcal{F}\{f\}[k]\}[n]\label{Eq:Henrici},
\end{align}
where one should note that if $g \in \mathbb{R}^N$, than Henrici's output is purely real-valued (see Section 1.C). Also note that in Marple's original work, only even-length signals were considered but it provided a conceptual layout to apply to odd-length signals. The cases structure of the Marple implementation may be re-written as:
\begin{align}
    \mathcal{H}\{f\}[n] ~=&~ \text{Im}\{\mathcal{F}^{-1}\{\text{(sgn}[k] + 1)\mathcal{F}\{f\}[k]\}\}\nonumber\\
    ~=&~ \text{Im}\{\mathcal{F}^{-1}\{\text{sgn}[k]\mathcal{F}\{f\}[k]\}\} + \underbrace{\text{Im}\{\mathcal{F}^{-1}\{\mathcal{F}\{f\}[k]\}\}}_{=\text{Im}\{f\}=0,~ \text{if}~ g\in\mathbb{R}^N}\nonumber\\
    ~=&~ \text{Re}\{\mathcal{F}^{-1}\{-i\, \text{sgn}[k] \mathcal{F}\{f\}[k]\}[n]\}\label{Eq:Marple_to_Henrici},
\end{align}
where we have used the fact that $\text{Im}\{y\} = \text{Im}\{y' + i~y''\} = \text{Re}\{-i~y\}~=~y''$, where $y$ is a complex-valued entity (with $y'$ and $y''$ as its real and imaginary parts, respectively).

For an even-valued N, the original form of the Marple implementation is utilized:
\begin{align}
    \text{Marple}: \mathcal{H}\{f\}[n] = \text{Im}\left\{\mathcal{F}^{-1}\left.\begin{cases}
    0 & k < 0\\
    \mathcal{F}\{f\}[k] & k=0\\
    2\mathcal{F}\{f\}[k] & 1\leq k < \frac{N}{2}\\
    \mathcal{F}\{f\}[k] & k=\frac{N}{2}
    \end{cases}\right\}\right\}\label{Eq:Marple_even}.
\end{align}
Similar to as in Eq. \ref{Eq:Marple_to_Henrici}, we can re-write this implementation as:
\begin{align}
    \mathcal{H}\{f\}[n] ~=&~ \text{Im}\{\mathcal{F}^{-1}\{\left(\text{sgn}[k] + 1 - \delta\left[\frac{N}{2}\right]\right)\mathcal{F}\{f\}[k]\}\}\nonumber\\
    ~=&~ \text{Im}\{\mathcal{F}^{-1}\{\text{sgn}[k]\mathcal{F}\{f\}[k]\}\} + \underbrace{\text{Im}\{\mathcal{F}^{-1}\{\mathcal{F}\{f\}[k]\}\}}_{=\text{Im}\{f\}=0,~ \text{if}~ g\in\mathbb{R}^N} - \underbrace{\text{Im}\{\underbrace{\mathcal{F}^{-1}\{\delta\left[\frac{N}{2}\right]\mathcal{F}\{f\}\}\}}_{\in \mathbb{R}^N,~ \text{if}~ g\in\mathbb{R}^N}}_{=0}\nonumber\\
    ~=&~ \text{Re}\{\mathcal{F}^{-1}\{-i\, \text{sgn}[k] \mathcal{F}\{f\}[k]\}[n]\}\label{Eq:Marple_to_Henrici_even};
\end{align}
thus, for even signals, the Marple and Henrici implementations are equivalent. It should be noted that in our numerical work, we find the difference between the results on the order of machine epsilon for double floating point arithmetic or exactly 0.0 in some cases.

\subsection{The Fourier transform of the Hilbert transformed signal}
\begin{align}
    \left | \mathcal{F}\{\mathcal{H}\{f\}\}[k]\right |^2 ~=&~ \left | -j\text{sgn}[k]\mathcal{F}\{f\}[k]\}\right |^2 = (\vec{1}-\delta[k])|\mathcal{F}\{f\}[k]\}|^2,
\end{align}
where $\vec{1}$ is a 1-vector of length N. Note that this finding disagrees with that of \cite{Poularikas1999} in which the square of the sign-function was implied to be a constant unity.

\subsection{For a real-valued input, the DHT output is real-valued}
\begin{align}
    \mathcal{H}\{f\}[n] ~=&~ \sum_k -i~\text{sgn}[k]~e^{i\frac{2\pi}{N}kn} \frac{1}{N}\sum_{n'} f[n]~e^{-i\frac{2\pi}{N}kn'}\nonumber\\
    ~=&~ \sum_k -i~\text{sgn}[k]~e^{i\frac{2\pi}{N}kn} \left[ F_\mathrm{even}[k] + i~F_\mathrm{odd}[k]\right],
\end{align}
where we use $F_\mathrm{even}$ and $F_\mathrm{odd}$ to represent the DFT of the even and odd components of the input signal $g$. Importantly, the DFT has symmetry properties such that $F_\mathrm{even}$ and $F_\mathrm{odd}$ have even and odd symmetry, respectively \cite{DFT}.

Assuming that $N$ is odd-valued such that summations span the same range negatively and positively, using symmetry of the signal and the DFT:
\begin{align}
    \mathcal{H}\{f\}[n] ~=~ \sum_k &-i~\text{sgn}[k]~\left[\cos\left(\frac{2\pi}{N}kn\right) + i~ \sin\left(\frac{2\pi}{N}kn\right)\right]\left[ F_\mathrm{even}[k] + i~F_\mathrm{odd}[k]\right]\nonumber\\
    ~=~ \sum_k &\left\{-i~\text{sgn}[k]\cos\left(\frac{2\pi}{N}kn\right)F_\mathrm{even}[k]\right.\nonumber\\ 
    &+~\text{sgn}[k]\sin\left(\frac{2\pi}{N}kn\right)F_\mathrm{even}[k]\nonumber\\
    &+~\text{sgn}[k]\cos\left(\frac{2\pi}{N}kn\right)F_\mathrm{odd}[k]\nonumber\\
    &\left.+~i~\text{sgn}[k]\sin\left(\frac{2\pi}{N}kn\right)F_\mathrm{odd}[k] \right\}\nonumber\\
    ~=~ \sum_k &\text{sgn}[k]\sin\left(\frac{2\pi}{N}kn\right)F_\mathrm{even}[k] +\text{sgn}[k]\cos\left(\frac{2\pi}{N}kn\right)F_\mathrm{odd}\label{Eq:Real_in_out},
\end{align}
which is real-valued.

\subsection{An even (odd) signal input to the DHT produces an odd (even) symmetry output}

Using Eq. \ref{Eq:Real_in_out} and comparing $\mathcal{H}\{f\}[n]$ and $\mathcal{H}\{f\}[-n]$ as applied to a signal with even-symmetry $f_\mathrm{even}$:
\begin{align}
    \mathcal{H}\{f_\mathrm{even}\}[n] ~=&~ \sum_k \text{sgn}[k]\sin\left(\frac{2\pi}{N}kn\right)F_\mathrm{even}[k]\\
    \mathcal{H}\{f_\mathrm{even}\}[-n] ~=&~ \sum_k -\text{sgn}[k]\sin\left(\frac{2\pi}{N}kn\right)F_\mathrm{even}[k]\\
    \therefore~\mathcal{H}\{f_\mathrm{even}\}[n] ~=&~ -\mathcal{H}\{f_\mathrm{even}\}[-n];
\end{align}
thus an even input generates an odd-symmetry DHT output.

Similarly for an odd-symmetry signal $f_\mathrm{odd}$:
\begin{align}
    \mathcal{H}\{f_\mathrm{odd}\}[n] ~=&~ \sum_k \text{sgn}[k]\cos\left(\frac{2\pi}{N}kn\right)F_\mathrm{odd}[k]\\
    \mathcal{H}\{f_\mathrm{odd}\}[-n] ~=&~ \sum_k \text{sgn}[k]\cos\left(\frac{2\pi}{N}kn\right)F_\mathrm{even}[k]\\
    \therefore~\mathcal{H}\{f_\mathrm{odd}\}[n] ~=&~ \mathcal{H}\{f_\mathrm{odd}\}[-n];
\end{align}
thus an odd input generates an even-symmetry DHT output.

\section{Simulated spectra}
\subsection{Gaussian input spectra}
In the main text of the manuscript, input spectra were of a Dawson form with a peaked lineshape (Gaussian) Hilbert transform. In this section, we consider the converse scenario, when input data is assumed to be Gaussian. The generation and constraints of the synthetic training and test data was identical to that of the main text. Similar to the main text results, the LeDHT is superior 98.74 \% and shown in Fig. S1.
\begin{figure}[h]
\centering
\includegraphics[]{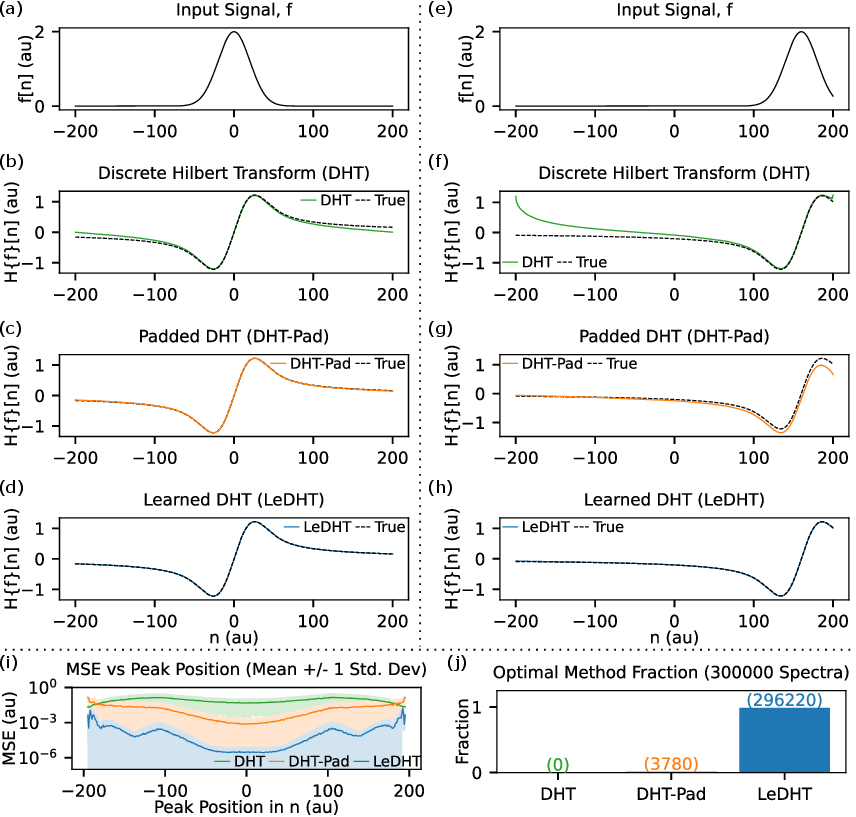}
\caption{(a,e) Example Gaussian input spectra. (b,f) DHT results showing significant errors. (c,g) Results using the DHT with extrema-value padding (DHT-Pad). (d,h) Results using the LeDHT. (i) Mean MSE ($\pm$ 1 standard deviation) calculated for the 300\thinspace000 stochastically generated test spectra as a function of peak center position. (j) Histogram of optimal method based on minimum MSE for test spectra dataset. Y-axis shows fraction, parenthetical values are number of spectra.}
\label{Fig:Synth_Peaks}
\end{figure}

\subsection{Multi-peak spectra}
Similar to above, we demonstrate the scenario with multi-peak input spectra composed of Gaussian lineshapes. The generation and constraints of the synthetic training and test data was identical to that of the main text. In the main text the LeDHT was always superior for 30\thinspace000 spectra, and under this new scenario the LeDHT is superior to either the DHT or DHT-Pad 99.57 \% of the 30\thinspace000 test spectra (see Fig. S2).
\begin{figure}[h]
\centering
\includegraphics[]{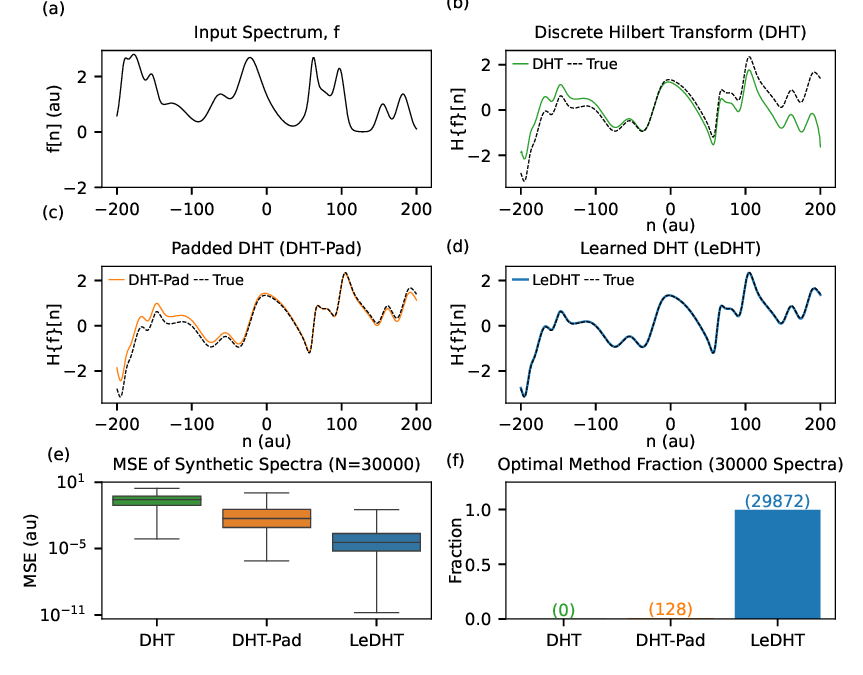}
\caption{(a) Example synthetic spectrum composed of 13 Gaussian functions. (b,c,d) The true Hilbert transform and the calculated Hilbert transforms using DHT, the DHT applied to extrema-value padded spectra (DHT-Pad), and the LeDHT, respectively. (e) Box plots of MSE for each method. Note the whiskers displays the extreme range of values. (f) Histogram of the optimal method (based on minimum MSE) for 30\thinspace000 test spectra. Y-axis shows fraction, parenthetical values are number of spectra.}
\label{Fig:Synth_Spectra}
\end{figure}

\subsection{Training on Gaussian-Dawson lineshape pairs and testing on Lorentzian-dispersive lineshape pairs}
In the main text, the lineshape types for training and testing were the same. In this section, we consider training with one type and testing with another. Ideally, the LeDHT matrix describes a relationship -- a transform -- and is not strictly a database for fitting input spectra. Thus, one would expect the LeDHT to still work, though possibly sub-optimally. Surprisingly, the LeDHT is superior to the DHT and DHT-Pad for all 300\thinspace000 test spectra (see Fig. S3).
\begin{figure}[h]
\centering
\includegraphics[]{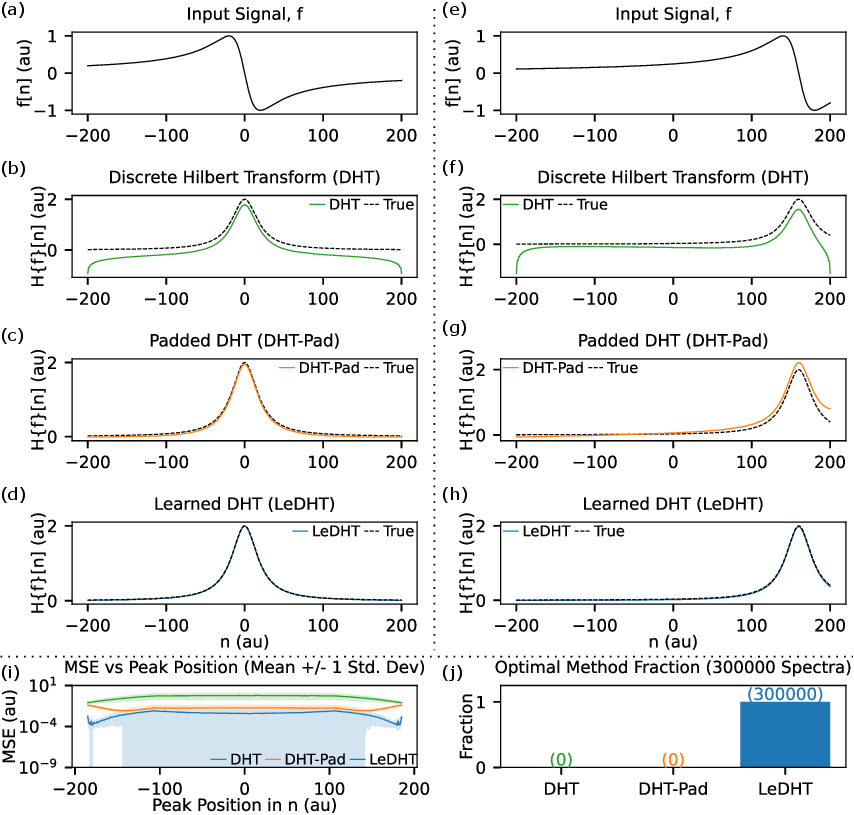}
\caption{(a,e) Example dispersive lineshape input spectra. (b,f) DHT results showing significant errors. (c,g) Results using the DHT with extrema-value padding (DHT-Pad). (d,h) Results using the LeDHT. The LeDHT was trained with Gaussian and Dawson lineshapes though the new input data is dispersive with a Lorentzian Hilbert transform. (i) Mean MSE ($\pm$ 1 standard deviation) calculated for the 300\thinspace000 stochastically generated test spectra as a function of peak center position. (j) Histogram of optimal method based on minimum MSE for test spectra dataset. Y-axis shows fraction, parenthetical values are number of spectra.}
\label{Fig:TrainGaussi_TestLorentz}
\end{figure}

\clearpage

\bibliography{SI_biblio}